\documentclass[epj]{svjour}

\usepackage{graphicx}
\usepackage{amsmath}

\begin{document}
\title{Autoionization of an ultracold Rydberg gas through resonant dipole coupling}
\author{T. Amthor\inst{1}\fnmsep\thanks{\emph{Present address:} Physikalisches Institut, Universit\"at Heidelberg, Philosophenweg 12, 69120 Heidelberg, Germany} \and J. Denskat\inst{1} \and C. Giese\inst{1} \and N. N. Bezuglov\inst{2} \and A. Ekers\inst{3} \and L. Cederbaum\inst{4} \and M. Weidem{\"u}ller\inst{1}\fnmsep$^\mathrm{a}$
%
}                     
%
%
\institute{Physikalisches Institut, Universit{\"a}t Freiburg, Hermann-Herder-Str. 3, 79104 Freiburg, Germany \and St. Petersburg State University, physical faculty, Institute of Physics, 198904 St. Petersburg, Russia \and University of Latvia, Laser Centre, LV-1002 Riga, Latvia \and Theoretische Chemie, Physikalisch-Chemisches Institut, Universita¨t Heidelberg, Im Neuenheimer Feld 229, 69120 Heidelberg, Germany}
\date{Received: date / Revised version: date}
%
\abstract{
We investigate a possible mechanism for the autoionization of ultracold Rydberg gases, based on the resonant coupling of Rydberg pair states to the ionization continuum.
Unlike an atomic collision where the wave functions begin to overlap,
the mechanism considered here involves only the long-range dipole
interaction and is in principle possible in a static system.
It is related to the process of intermolecular Coulombic decay (ICD).
In addition, we include the interaction-induced motion of the atoms and the effect of multi-particle systems in this work. We find that the probability for this ionization mechanism can be increased in many-particle systems featuring attractive or repulsive van der Waals interactions. However, the rates for ionization through resonant dipole coupling are very low. It is thus unlikely that this process contributes to the autoionization of Rydberg gases in the form presented here, but it may still act as a trigger for secondary ionization
processes.
As our picture involves only binary interactions, it remains to be investigated if collective effects of an 
ensemble of atoms can significantly influence the ionization probability.
Nevertheless our calculations may serve as a starting point for the investigation of more complex systems, such as the coupling of many pair states proposed in \cite{tanner2008}.
\PACS{
	{32.80.Zb}{Autoionization} \and
	{32.80.Ee}{Rydberg states} \and
	{34.20.Cf}{Interatomic potentials and forces}
     } 
} 
\maketitle

\section{Introduction}

Ultracold Rydberg gases have been found to autoionize, typically on time scales of $\mu$s
\cite{robinson2000}.
Different processes have been identified as possible mechanisms leading to this ionization.
Direct photoionization by black-body radiation is one of the important processes.
However, under the typical conditions of ultracold experiments, the photoionization rates are still dominated by interaction-induced collisions of Rydberg atoms, which lead to Penning ionization,
\begin{equation}
	Ryd(n) + Ryd(n) \rightarrow Ryd(n') + ion + e^- \; .
\end{equation}
In this process the long-range interactions accelerate the initially ultracold Rydberg atoms
towards each other until they collide, and the ionization cross sections are typically very large
\cite{olson1979,robicheaux2005}.
\begin{figure}
 \begin{center}
  \includegraphics[scale=1]{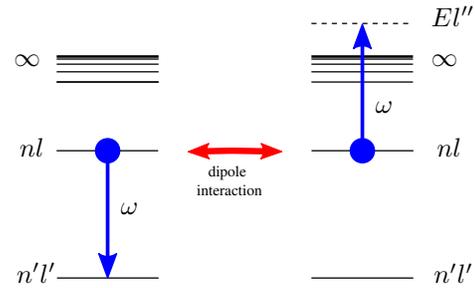}
 \end{center}
  \caption{Resonant coupling to the continuum. Two atoms are prepared in state $nl$ at distance $R$. Due to dipole coupling, one atom can be ionized, while the other one is transferred to $n'l'$ with ${n^*}'<{n^*}/\sqrt{2}$.}
  \label{fig:CouplingScheme}
\end{figure}
Recent experiments have identified collisional ionization to be one of the main ionization mechanisms of ultracold Rydberg gases \cite{li2005,amthor2007}, although other processes also
seem to be involved in the dynamics of Rydberg autoionization. Experiments with dense samples
show that ions can be observed even before collisions may have occurred \cite{tanner2008}.
This is attributed to a series of resonant dipole-coupled pair states which may allow one of the atoms of a pair to ionize within a time much shorter than the timescale of interaction-induced motion. In more dilute gases, simulations of atom trajectories in many-particle systems are found to underestimate the observed ionization rate at high principal quantum
numbers \cite{amthor2007b}. Apparently there exists a distance-dependent coupling of
two-particle or many-particle states to other states involving the continuum.
This dipole coupling induced autoionization may also be relevant for the
stability of Rydberg macromolecules \cite{flannery2005}.

Here, we consider the simple case of a direct coupling of a pair state to the ionization
continuum (see Fig.~\ref{fig:CouplingScheme}).
We restrict ourselves to the range of distances where the wavefunctions do not yet overlap.
Once the atoms reach the overlap region, the ionization probability can be enhanced by orders of
magnitude \cite{averbukh2004}.
Two atoms, both in state $|nl\rangle$ and separated in space by a distance $R$,
are coupled by the dipole-dipole interaction in such a way that one atom is
ionized, while the other one undergoes a transition to a lower Rydberg state $|n'l'\rangle$.
Such a process becomes possible whenever the energy difference between the
$|nl\rangle$ and $|n'l'\rangle$ states is larger than (or at least equal to) the binding energy
of the Rydberg electron in the $|nl\rangle$ state. The rate of this process depends strongly
on the initial Rydberg state and on the distance between the atoms.
The process has been discussed for thermal gases \cite{katsuura1965} and for dense
cold gases \cite{hahn2000}.
The mechanism is related to the interatomic or intermolecular Coulombic decay (ICD),
a collective autoionization process which has been intensely studied in clusters \cite{zantra2000,cederbaum1997,marburger2003b,jahnke2004},
and has recently been observed between OH$^-$ and H$_2$O molecules \cite{aziz2008}.
ICD occurs when an excited particle relaxes into a lower state, while in a radiationless process a neighboring particle picks up the energy and is ionized.
While in the typical ICD process one of the involved particles is an ion (see references above) or atom \cite{barth2005} in an excited state with only one or a few
empty low-lying energy levels,
highly excited Rydberg atoms feature a large number of unpopulated states below
the Rydberg state.
In the present work we investigate how relevant such a process can be
for the autoionization of a dilute ultracold Rydberg gas, when interaction-induced
motion of the atoms and many-particle effects are taken into account.

The paper is organized as follows:
In Section \ref{sec:Calculation} the calculation of the ionization rates based on dipole
interaction is described. The ionization probability of a pair of atoms on an attractive
interaction potential is discussed in Section \ref{sec:PairMotion}.
Section \ref{sec:Surrounding} addresses the influence of the surrounding atoms in a cloud on
this accelerated pair.
Finally, autoionization in a many-body system of repulsively interacting atoms is investigated in Section \ref{sec:Repulsive}.

\section{Calculation of rates}
\label{sec:Calculation}

\begin{figure}
  \includegraphics[scale=1]{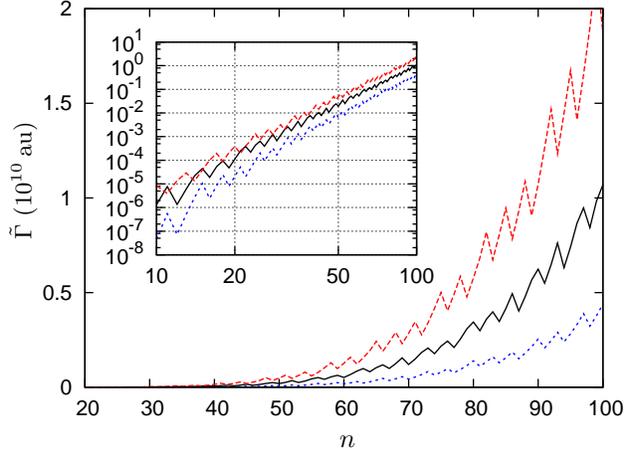}
  \caption{Rate coefficient $\tilde\Gamma$ for different initial states $nl$ of rubidium
	(in atomic units): $n$s (dotted line), $n$p (solid line), $n$d (dashed line).
	The inset shows the same data in a log-log scale to visualize the power law scaling.}
  \label{fig:RatesVsN}
\end{figure}

\begin{figure}
  \includegraphics[scale=1]{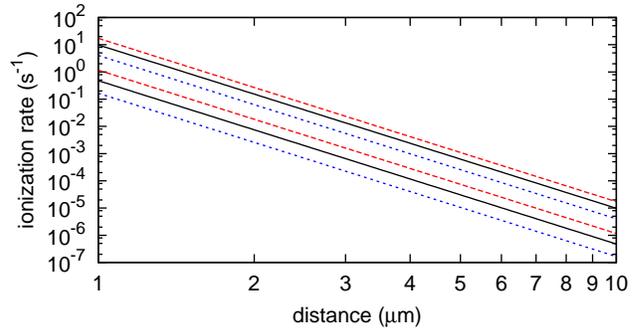}
  \caption{Distance dependence of the rates $\Gamma$ (given in $s^{-1}$)
	for different initial states $nl$:
	$n$s (dotted line), $n$p (solid line), $n$d (dashed line) for $n=100$ (upper
	three traces) and $n=60$ (lower three traces).}
  \label{fig:RatesVsR}
\end{figure}

Consider a pair of atoms, both in a state $|nl\rangle$, at an internuclear separation $R$.
The rate for the transition $|nl,nl\rangle \rightarrow |n'l',E,l''\rangle$ can be expressed
in terms of the dipole matrix element of the bound-bound transition and the
photoionization cross section for the bound-free transition \cite{galitskii1981}.
Atomic units are used in the following. Averaging over a uniform angular distribution, the rate
is given by
\begin{equation}
  \label{eq:gammaprime}
  \Gamma_{nl,n'l'}(R)=\frac{c\sigma ^{}_{PI}(nl)}{\pi\omega_{nl,n'l'}R^6} |D_{nl,n'l'}|^2\; ,
\end{equation}
where $D_{nl,n'l'}$ is the reduced matrix element for the electric dipole transition between the two bound states $|nl\rangle$ and $|n'l'\rangle$, $\omega_{nl,n'l'}>0$ denotes the energy difference between these states, and $\sigma_{PI}$ is the photoionization cross section for a transition from $|nl\rangle$ to the continuum with the same energy difference.

The total ionization rate must take into account transitions to all lower states $|n'l'\rangle$ of one atom, which yield sufficient energy for the ionization of the other atom. The ionization energy of an atom in the state $|nl\rangle$ with the effective quantum number $n^*$ is equal to its binding energy, $E_{ion}=|E_{n^*}|=1/(2n^{*2})$. This means that ionization is possible for transitions to levels ${n'}^*$ with binding energies $|E_{{n'}^*}|=1/(2({{n'}^*})^2)$ larger than the threshold value $|E_{th}|=2E_{ion}=1/n^{*2}$. From here we immediately obtain the condition for ionizing levels as  ${n'}^* \leq n^*/\sqrt{2}$. The total rate can then be obtained by summing over all states $n'^{*}$ that satisfy the above ionization condition:
\begin{equation}
  \Gamma_{nl}(R)=\sum_{\substack{{n'}^*\leq {n^*}/\sqrt{2}\\l'=l\pm 1}}\Gamma_{nl,n'l'}(R) \;,
\end{equation}

The photoionization cross sections are evaluated using the semiclassical formula given in
\cite{bezuglov1999,miculis2005,beterov2007},
\begin{equation}
  \label{eq:photo1}%
  \sigma_{PI}(n,l)=\frac{2}{3c}\frac{1}{\omega_{nl,n'l'}n^{*\, 3}}\sum\limits_{\Delta l=\pm 1}
\frac{L_{c}^{5}}{l \! + \! 0.5} D^{2}_{\Delta l}(\omega _{nl,n'l'} ) \, ,
\end{equation}
where the sum corresponds to the transitions $l \rightarrow l''=l+\Delta l$ into the continuum with $L_{c}=\mathrm{max}\{ l,l'' \}$. The coefficients 
\begin{equation}
\label{eq:photo2}%
D_{\Delta l}= \left[ -\frac{\sin
(\pi \Delta \mu )}2-\frac{\sqrt{\pi }}x \, \Phi _{\Delta
\mu}^{^{\prime }} (x)+\Delta l_q \sqrt{\frac \pi x} \, \Phi
_{\Delta \mu}(x)\right]
\end{equation}
\noindent are expressed via the conventional Airy function and its derivative:
\begin{align}
\label{eq:airy}%
\Phi _{\Delta \mu}(x)&=\frac {1}{\sqrt{\pi }} \int\limits_0^\infty
d\xi \; \cos (x\xi +\xi ^3/3+\pi \Delta \mu ) \\
\quad  \Phi _{\Delta \mu} ^{^{\prime }}(x)&=\frac d{dx}\Phi
_{\Delta \mu}(x)  \; ; \quad x=\left( \frac{\omega
^{}_{nl,n'l'} L_{c}^{3}}{2}\right) ^{2/3} \\%
\quad \Delta l_q&=\Delta l+\frac{1}{5L_{c}}\left(1-\frac{1}{n^{*2}\,\omega^{}_{nl,n'l'}} \right) \, . 
\end{align}
Here $\Delta \mu$ denotes the difference between the quantum defects $\mu_l$ for series $l$ and $l''$ involved into the photoionization: $\Delta \mu=\mu_{l+\Delta l}-\mu_l$. For small and large values of the argument $x$ the functions $\Phi _{\Delta \mu}(x)$ and $\Phi _{\Delta \mu} ^{^{\prime }}(x)$ can be approximated by simple asymptotical expressions \cite{bezuglov1999}.

The radial matrix elements for the transitions to bound states are calculated numerically by integrating the Schr\"o\-dinger equation for the corresponding energies using the Numerov algorithm \cite{zimmerman1979}.
For the low-lying levels the wavefunctions are determined using a model potential fitted to one-electron energies \cite{marinescu1994}.

In order to eliminate the $R$ dependence of the rate, we define a coefficient $\tilde\Gamma_{nl}$, such that
\begin{equation}
 \label{eq:gammatilde}
  \Gamma_{nl}(R)=\frac{\tilde\Gamma_{nl}}{R^6} \; .
\end{equation}
The calculated coefficient $\tilde\Gamma_{nl}$ for rubidium is plotted in
Fig.~\ref{fig:RatesVsN} as a function of the
quantum numbers $n$ and $l$.
The oscillatory behavior of the rate coefficient has the following simple explanation. As discussed above, the ionization can occur if the binding energy $E_{n'^*}$ of the final level $n'^*$ exceeds the threshold value $E_{th}$. The excess energy $\varepsilon=|E_{{n'}^*}|-|E_{th}|=1/2n'^{*2}-1/n^{*2}$ is carried away by the emitted electron. Since the photoionization cross section decreases rapidly with increasing energy of the photoelectron $\varepsilon$, the transitions to the $|n'l'\rangle$ level closest to the threshold will give the main contribution to the ionization rate. The energy spacing between the bound levels increase with decreasing $n$. Therefore for different $|nl\rangle$ levels the corresponding highest possible ionizing $|n'l'\rangle$ levels 
appear at different separations from the threshold $E_{th}$,
leading to different excess energies $\varepsilon$.
Hence, the ionization rates for different $|nl\rangle$ levels will be larger or smaller, depending on how close to the threshold is the first ionizing $|n'l'\rangle$ level. 

\begin{table}
 \caption{Fit parameters for a power law scaling of the rate coefficient of the form
	$\tilde\Gamma=a n^b$.}
 \label{tab:FitParameters}
 \begin{center}
 \begin{tabular*}{0.3\textwidth}{@{\extracolsep{\fill}}ccc}
  \hline \hline
    $l$ & $a$ & $b$ \\
    \hline
    0 (s)  & 0.014 & 5.74 \\
    1 (p)  & 0.092 & 5.52 \\
    2 (d)  & 0.240 & 5.46 \\
  \hline \hline
 \end{tabular*}
 \end{center}
\end{table}
Neglecting the oscillations, the average rate coefficient exhibits a power law scaling with the principal
quantum number, $\tilde\Gamma=a n^b$, as can also be seen from the logarithmic
representation in Fig.~\ref{fig:RatesVsN} (inset).
The best fit parameters $a$ and $b$ corresponding to the different quantum numbers $l$
are listed in Table \ref{tab:FitParameters}.
Note that the scaling exponent $b$ can be calculated analytically for the case of hydrogen
\cite{bezuglov1995}, yielding $b=5.333$ independent of $l$, which is close to the fitted value for rubidium with $l=2$.

Figure~\ref{fig:RatesVsR} shows the calculated rates $\Gamma_{nl}$ as a function
of the interatomic distance $R$ for different quantum numbers.
For typical interatomic distances and high principal quantum numbers the rates
are in the mHz range, so that the contribution of this ionization process seems
negligible for experiments with Rydberg gases.
Rates above 100\,Hz are only possible for very high principal quantum numbers
($n>100$) at short distances ($\sim 1\,\mu$m).
Due to the excitation blockade observed for high Rydberg states, the minimum distance between Rydberg atoms shortly after the excitation is limited, typically to a few $\mu$m \cite{tong2004,singer2004}.
Furthermore, the interaction-induced acceleration will not allow the atoms to remain at
rest on the timescale of an autoionization process. This phenomenon will be discussed in the next section.

Despite the low rates, the process described here can still act as a trigger for additional ionization processes. Collisions of other Rydberg atoms with the electron and ion produced can lead to secondary 
ionization and ionization avalanches. In some configurations, these avalanches are even found to create
ultracold plasmas \cite{li2004}.

\section{Ionization probability during motion}
\label{sec:PairMotion}

Two interacting Rydberg atoms will experience a force given by the gradient of the interaction potential, which can be either attractive or repulsive.
Consider an atom pair on an attractive van der Waals potential with the
coefficient $C_6$ such that
\begin{equation}
 \label{eq:vdWPotential}
 V_{vdW} = - \frac{C_6}{R^6} \; .
\end{equation}
The atoms will attract each other and collide within several $\mu$s for the typical inital distance of a few $\mu$m. Due to the motion of the atoms the resonant ionization rates become time-dependent, increasing dramatically as atoms approach each other.

In order to investigate the importance of this mechanism in such a dynamic system, we calculate the ionization probability during the motion. Let $p(t)$ be the probability for having been
ionized at a time $t$, and $\mathcal P=p(t_{coll})$ be the probability for ionization before a
collision occurs. $\mathcal P$ can be evaluated by integration,
\begin{eqnarray}
 dp &=& (1-p(t)) \, \Gamma(R(t)) \, dt\\
 \label{eq:Pcoll}
 \mathcal P &=& 1- \exp\left(-\int_0^{t_{coll}} \Gamma(R(t)) \, dt\right) \; .
\end{eqnarray}
As long as the exponent in Eq.~(\ref{eq:Pcoll}) is small compared to unity,
this can be simplified as
\begin{equation}
 \mathcal P \approx \int_{0}^{t_{coll}} \Gamma(R(t)) dt \; ,
 \label{eq:probMotion}
\end{equation}
where $t_{coll}$ is the time at which the pair distance is so small that the electronic wave functions begin to overlap. This distance is set to $R_{coll}\simeq 4n^{*2}$, below
which collisional ionization occurs \cite{robicheaux2005}.
$R(t)$ is evaluated using the classical equation of motion on the potential given by Eq.~(\ref{eq:vdWPotential}). The coefficient $C_6$ is taken from the calculations in
Ref.~\cite{singer2005b}.

The collision time $t_{coll}$ for atoms with mass $M$ initially at a distance $R_0$
may be evaluated analytically assuming that the initial velocity of the atoms is zero.
In this case the collision occurs with zero impact parameter and 
$t_{coll}$ can be estimated by writing down the 
equation of energy conservation with the reduced mass $\mu=M/2$,
\begin{equation}
  V(R_0) = V(R)+\frac{1}{2}\mu\left(\frac{dR}{dt}\right)^2 \; ,
\end{equation}
and integrating from $R=R_0$ to the collision radius $R=R_{coll}$,
\begin{equation}
  \label{eq:intcoll}
  \int_0^{t_{coll}} dt = \int_{R_0}^{R_{coll}} \frac{dR}{ \sqrt{ 2\mu^{-1}\left(V(R_0)-V(R)\right) }} \; .
\end{equation}
The collision time does not depend strongly on the exact value of $R_{coll}$ and can be estimated
by setting $R_{coll}=0$.
It is also independent of the exact shape of the interaction potential at short distances, as the atoms spend most of their time at larger separations. The integral can thus be solved for a simple van der Waals potential to yield the collision time 
\begin{equation}
  \label{eq:collisiontimevdw}
  t_{coll}\approx 0.22\sqrt{\frac{M}{C_6}}R_0^4 \; .
\end{equation}
For large initial distances (i.e. $R_{0}/R_{coll}\gg 1$, $V(R_0)=0$),
the ionization probability can be written analytically as
\begin{equation}
 \mathcal P\approx \sqrt{\frac M{4C_6}}\tilde \Gamma \int_{R_{coll}}^{R_{0}}dRR^{-3}\approx \sqrt{\frac M{4C_6}}\tilde \Gamma \frac 1{2R_{coll}^2} \; .
 \label{eq:probMotion2}
\end{equation}

\begin{figure}
  \begin{center}
   \includegraphics[scale=1]{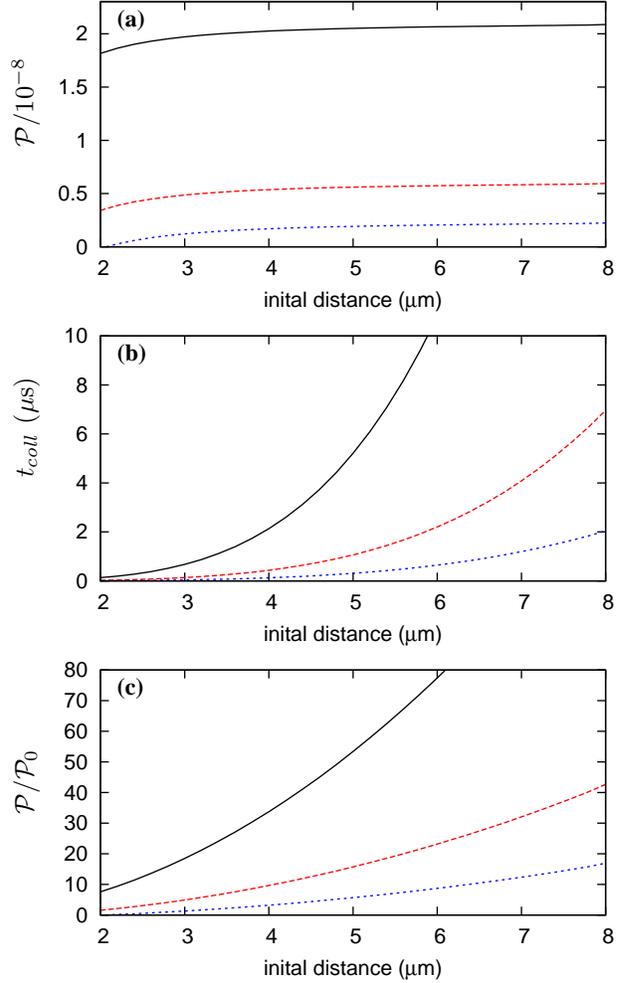}
  \end{center}
  \caption{Various parameters of ionizing Rydberg pairs 60D--60D (solid lines),
	80D--80D (dashed lines), 100D--100D (dotted lines) as a function of the initial
	interatomic separation $R_0$:
	(a) resonant dipole ionization probability $\mathcal P$ before the collision;
	(b) collision time $t_{coll}$;
	(c) relative increase of the ionization probability $\mathcal P/\mathcal P_0$
	for a moving pair compared to a static pair.}
  \label{fig:Motion}
\end{figure}

Fig.~\ref{fig:Motion}(a) shows the ionization probability $\mathcal P$, calculated by numerical integration of Eq.~(\ref{eq:probMotion}), as a function of the initial distance for pairs of atoms in the states 60D ($C_6=1\times 10^{21}$), 80D
(red, $C_6=2.4\times 10^{22}$), and 100D (red, $C_6=2.8\times 10^{23}$).
The ionization probabilities increase slightly with increasing initial distance, approaching the
value found from Eq.~\ref{eq:probMotion2}.
In order to reach these probabilities, the observation time must be at least as long as the collision time plotted in Fig.~\ref{fig:Motion}(b). For a limited observation time the rates will decrease again with increasing $R_0$.
Despite the higher value of $\tilde\Gamma$ for increasing principal quantum number $n$, the
probability $\mathcal P$ decreases, as the atoms spend only a short time in close proximity
to each other before they collide. If we let
\begin{equation}
 \label{eq:pzero}
 \mathcal P_0 = \Gamma(R(0)) t_{coll} \approx 0.22\sqrt{\frac M{C_6}}\frac{\tilde \Gamma }{R_0^2}
\end{equation}
be the ionization probability for a static pair interacting during the same time $t_{coll}$,
then we obtain the relative probability $\mathcal P/\mathcal P_0 $ (see
Fig.~\ref{fig:Motion}(c)). For large $R_0$ we obtain from Eqs. (\ref{eq:probMotion2})
and (\ref{eq:pzero}) the relation
 \begin{equation}
\mathcal P/\mathcal P_0 \approx 1.14\,R_0^2/R_{coll}^2 \; .
 \label{eq:relative}
\end{equation}
For typical initial distances $R_0$ the attractive motion increases the
ionization probability by one to two orders of magnitude compared to the static pair.

\section{Influence of surrounding atoms}
\label{sec:Surrounding}

So far we have considered the simple case of two atoms.
In a large ensemble of atoms, e.g., an ultracold atom cloud,
all the surrounding atoms will contribute to the ionization rate.
But as the arrangement of atoms is random and not a crystalline structure,
there will always be exactly one nearest neighbor.
Consider an atom with its nearest neighbor at a distance $R_0$. According to
Eq.~(\ref{eq:gammatilde}), the ionization rate for this pair is given by
\begin{equation}
 \Gamma_{pair} = \frac{\tilde\Gamma}{R_0^6} \; .
\end{equation}
To estimate the importance of the surrounding atoms, we assume a constant density $\rho$ and calculate an additional rate $\Gamma_{s}$ induced by all other atoms except the nearest neighbor
by integrating from the radius $R_0$ to infinity (as the nearest neighbor is at $R_0$ by definition, all other atoms must be at larger distances):
\begin{equation}
 \Gamma_{s} = \int_{R_0}^\infty 4\pi R^2 \, \rho \,\, \Gamma (R) dR 
 \label{eq:gammaS}
\end{equation}
Here $\Gamma(R)=\tilde\Gamma/R^6$ is the ionization rate of a pair at distance $R$.
The integral yields
\begin{equation}
 \Gamma_{s} =\frac{4\pi}{3}\rho\tilde\Gamma\frac{1}{R_0^3} \; .
\end{equation}
For a homogeneous density $\rho$ the average nearest neighbor distance can be estimated as $a=(5/9)\rho^{-1/3}$ \cite{hertz1909}.
Assuming $R_0=a$, we can express $\rho$ in terms of $R_0$ and obtain
\begin{eqnarray}
 \Gamma_{s} &=& \frac{4\pi}{3}\left(\frac{5}{9}\right)^3\tilde\Gamma\frac{1}{R_0^6} \\
	  &\approx& 0.718\,\Gamma_{pair} \; .
\end{eqnarray}
The ionization rate induced by all the atoms with $R>a$ (only pairwise
interactions are considered) is therefore smaller than the rate induced only by the nearest neighbor at $R_0=a$. The ratio $\Gamma_s/\Gamma_{pair}$ will be even smaller for very close pairs with $R_0<a$.
If the atoms are accelerated due to an attractive van der Waals interaction, the
$1/R^6$ scaling of the potential accounts for a much larger acceleration towards the nearest neighbor compared to any of the surrounding atoms. 
The nearest neighbors will therefore approach each other quickly and
the influence of the surrounding atoms will be even smaller compared to a static system.\\

\begin{figure}
  \begin{center}
   \includegraphics[scale=1]{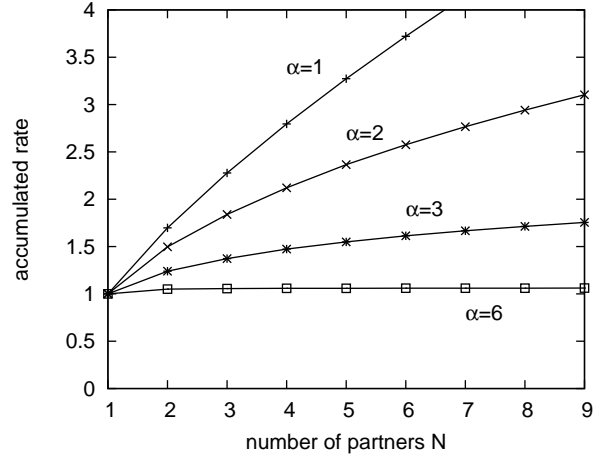}
  \end{center}
  \caption{Monte-Carlo simulation for the accumulated rates $\sum_{n=1}^N 1/R_n^\alpha$ when the $N$ nearest
	neighbors are taken into account and different scalings $\alpha=1,2,3,6$.
	While for $\alpha=6$ the rate approaches a limit as $N\rightarrow\infty$
	and is mainly determined by the first neighbor, the scalings $\alpha\le 3$ lead to a diverging rate
	where all surrounding neighbors have to be considered.
	All values are normalized to the corresponding $N=1$ rate. }
  \label{fig:AccRatesMC}
\end{figure}

In order to take the complete distribution of nearest neighbors into account,
a Monte Carlo simulation has been performed, where a large number of atoms are placed randomly in space.
Fig.~\ref{fig:AccRatesMC} shows the results for different scalings
$\Gamma\propto 1/R^\alpha$ of an accumlative quantity $\Gamma$
for a randomly picked atom, taking different numbers $N$ of its neigbors
into account. The results shown are the total rates of 100 randomly picked atoms,
scaled to the respective value of a single neighbor for easier comparison.
For a scaling of $\alpha=6$ clearly only the first neighbor has an effect, while for
$\alpha\leq 3$ the accumulated rate increases quickly.\\

Generally, for quantities $\Gamma$ scaling as $1/R^3$ the influence of the surrounding
atoms becomes significant.
This can be seen from the fact that the integral in Eq.~(\ref{eq:gammaS}) then diverges and can only be evaluated with an upper limit $R_{max}$. Expressing the limits of the integral in terms of the average nearest neighbor distance $a$ with $R_0=a$ and $R_{max}=\beta a$ ($\beta>1$) the integral yields
\begin{equation}
 \Gamma_{3,s} = 2.155\ln \beta \; \Gamma_{3,pair}
\end{equation}
where $\Gamma_{3,s}$ and $\Gamma_{3,pair}$ now represent arbitrary accumulative quantities scaling as $1/R^3$. For $\beta=100$ (a realistic value considering a magneto-optically trapped cloud), the contribution of the surrounding atoms outweighs the nearest neighbor by an order of magnitude.
Depending on the scaling of the quantity $\Gamma$ with the interatomic distance,
the influence of other atoms surrounding a nearest neighbor pair may or may not be significant.
This should be kept in mind when estimating energy shifts of an atom interacting via van der Waals ($1/R^6$) or dipole-dipole ($1/R^3$) interactions. In the van der Waals case, it may be sufficient to consider only the nearest neighbor, while the dipole-dipole interaction requires taking a larger number of atoms into account.
Note that in the case of dipole-dipole interactions, the angular dependence also has to be considered.

The estimates presented in this section justify the chosen approach to consider an attractively interacting
sample as a collection
of pairs, and to derive ionization probabilities in a pair picture as described above,
even if the atoms are embedded in a larger ensemble.\\

\section{Repulsive many-particle dynamics}
\label{sec:Repulsive}

\begin{figure}
  \includegraphics[scale=0.95]{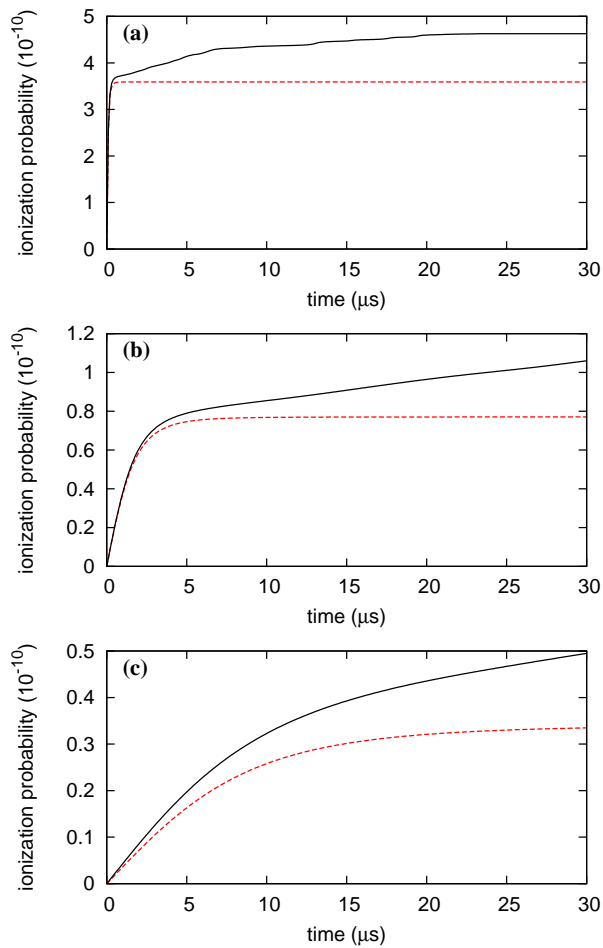}
  \caption{Ionization probability for pair of repelling atoms in state 60S ($C_6=-10^{21}\,$au)
	for different initial distances: (a) 2$\mu$m, (b) 4$\mu$m, (c) 6$\mu$m. The dashed
	lines show the calculated probabilities for two atoms in free space, the solid 
	lines show the results of multi-particle calculations for the same pair embedded in
	a cloud of interacting atoms.}
  \label{fig:RepulsiveRates}
\end{figure}
The derivation of the resonant ionization rates in terms of a two-atom picture is only
justified if the atoms are pulled together by an attractive force, thereby reducing the
influence of the surrounding cloud.
If the atoms are prepared in states exhibiting repulsive interaction, this assumption cannot be made. In this case the dynamics of the whole sample of atoms must be considered.
An isolated pair of atoms will be drawn apart immediately, so that the ionization probability is small.
In a many-particle system where all atoms are repelling each other, each particle may come into the vicinity of other atoms several times while moving across the cloud \cite{amthor2007b}. This increases the average time spent at close distance and thus increases the ionization probability.
In this simplified picture, the atoms do not collide at all. We thus need to specify the probability for resonant dipole ionization as a function of time, $\mathcal P=\mathcal P(t)$.

Figure~\ref{fig:RepulsiveRates} shows the ionization probability for a pair of atoms in the 60S state (with $C_6=-10^{21}\,$au). Two isolated atoms in free space will be accelerated away from each other, and the ionization probability reaches a steady-state value.
The same nearest-neighbor pair of repelling atoms placed in the middle of a large atom cloud
has a higher ionization probability, as both atoms come close to other particles while moving
away from each other.
The probability to ionize one of these two specific atoms increases steadily in time.
In the calculation presented in Fig.~\ref{fig:RepulsiveRates},
a density of Rydberg atoms of $1.5\times 10^9\,$cm$^{-3}$ is assumed, and the atoms surrounding the
initial pair are placed randomly in space. The dynamics of the system is then calculated
by solving the equations of motion for all atoms according to the van der Waals interaction
potentials in the same way as described in Ref. \cite{amthor2007b}, and the results are
averaged over 50 runs of the simulation.
For typical nearest-neighbor distances of about $4\,\mu$m, the surrounding atoms in this example enhance the probability to ionize within $30\,\mu$s by roughly a factor of 1.4.
However, the total ionization probability within this time is of the order of $10^{-10}$,
which is far from being observable experimentally.

\section{Conclusion}

We have presented calculations of ionization rates and estimates of many-particle effects for an
autoionization process of ultracold Rydberg atoms based on resonant dipole--dipole
interaction.
Our approach involves only binary interactions and a direct coupling to a continuum state.
The estimated rates for this process are not sufficient to explain the observed
autoionization in dense or strongly interacting Rydberg gases not caused by
collisions \cite{tanner2008,amthor2007b}.
Taking into account the motion and the influence of a many-particle environment, the expected
ionization probability for this process increases, but it is still too low to be
observed experimentally. This suggests that other interaction-induced ionization
processes exist, which contribute considerably during time scales shorter than the collision time of the atoms.
Furthermore, even a process with a low intrinsic rate may act as a trigger for secondary ionization.

We believe that collective effects of larger ensembles of interacting atoms may have a great influence on dipole coupling induced autoionization.
As a next step, the model could be extended to a coupling of many
intermediate pair states as proposed by \cite{tanner2008}, and for further analysis the system
could be described in terms of multi-particle states instead of pair states.

The discussion of the dynamics in many-body systems presented here is not restricted to the
autoionization process but may also be helpful for other dynamical processes involving
Rydberg interactions.

\section{Acknowledgements}

We acknowledge financial support by the Deutsche For\-schungs\-ge\-mein\-schaft
(grant no. WE2661/10-1), the EU FP6 TOK project LAMOL (contract MTKD-CT-2004-014228),
the Russian Foundation for Basic Research (grant no. 09-02-92428),
and the Latvian Science Council.
The authors thank S. Saliba for helpful discussions.


\end{document}